# The Quantum Compass Mechanism in Cryptochromes

Chengye Zou[1], Ya-jun Liu[1], and Beibei Wang[1,*]

[1] *Center for Advanced Materials Research, Beijing Normal University at Zhuhai, Zhuhai 519087, China;*

* *Corresponding authors: Beibei Wang: <u>bbwang@bnu.edu.cn</u>.*

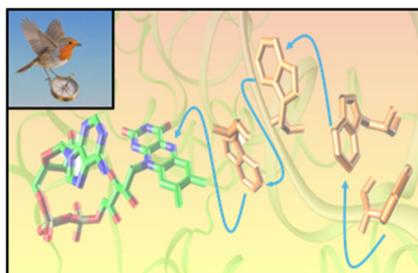

**Abstract:** Cryptochrome flavoproteins are prime candidates for mediating magnetic sensing in migratory animals via the radical pair mechanism (RPM), a spin-dependent process initiated by photoinduced electron transfer. The canonical FAD–tryptophan radical pair exhibits pronounced anisotropic hyperfine couplings, enabling sensitivity to geomagnetic fields. However, maintaining spin coherence under physiological conditions and explaining responses to weak radiofrequency fields remain unresolved challenges. Alternative radicals, such as superoxide ($O_2^{\bullet-}$) and ascorbate ($Asc^{\bullet-}$), have been proposed to enhance anisotropy or suppress decoherence. This review summarizes the quantum basis of magnetoreception, evaluates both canonical and alternative radical pair models, and discusses amplification strategies including triads, spin scavenging, and bystander radicals. Emphasis is placed on how molecular geometry, exchange and dipolar interactions, and hyperfine topology modulate magnetic sensitivity. Key open questions and future directions are outlined, highlighting the need for structural and dynamical data under physiological conditions.

**Keywords:** Radical pair mechanism, Cryptochrome, Spin dynamics, Magnetoreception.

## 1. INTRODUCTION

Humans have employed the compass as a directional tool for over two millennia, yet migratory birds possess an intrinsic ability to sense the direction of the Earth's geomagnetic field, enabling them to traverse vast distances along precise migratory routes. Two primary mechanisms have been proposed to account for the biological detection of such weak magnetic fields (ranging from 25 to 65 μT). One hypothesis posits the existence of ferromagnetic particle clusters located in the avian beak, functioning as biogenic magnetometers [1-3]. The alternative hypothesis is the radical pair mechanism (RPM). In 1978, Klaus Schulten and colleagues proposed that avian magnetoreception could be explained by a chemical reaction system in the retina based on radical pairs. This system is sensitive to magnetic fields because the geomagnetic field modulates the spin dynamics of the radicals, specifically the interconversion between singlet and triplet states, thereby influencing the yield of downstream reaction products [4]. In 2000, Thorsten Ritz and collaborators further postulated that this mechanism may be mediated by cryptochrome proteins located in the retina.5 Subsequent theoretical and experimental studies have focused on the cofactor flavin adenine dinucleotide (FAD) within cryptochromes, which can form the radical pair $FAD^{\bullet-} - Trp^{\bullet+}$ upon photoexcitation and electron transfer. Cryptochromes containing FAD, present in avian retinal cells, are now considered a leading candidate system for detecting the direction of magnetic fields [6-10]. Recent in vitro experiments have confirmed that radical pairs $FAD^{\bullet-} - Trp^{\bullet+}$ in certain migratory songbirds are indeed magnetically sensitive [11]. This article will provide a detailed overview of the relevant theories and research progress on the role of radical pairs in cryptochromes in sensing magnetic fields.

## 2. The Radical Pair Mechanism

In 1976, Schulten and colleagues studied the recombination dynamics of radical pairs composed of pyrene and 3,5-dimethoxy-N, N-dimethylaniline (DMDMA) in solution. They discovered that external magnetic fields could modulate the singlet and triplet yields of radical pair recombination products. This modulation was found to depend on both the magnetic field strength and the presence of hyperfine coupling within the radical pair. These observations were consistent with theoretical predictions that hyperfine-coupled radical pairs exhibit spin state interconversion sensitive to external magnetic environments [12]. In fact, the time evolution of the spin states of radical pairs is governed by the Hamiltonian $\hat{H}$, which incorporates hyperfine interactions, the Zeeman effect, and other relevant electronic contributions. The time evolution of spin states can be described by the stochastic Liouville equation [4], while the electronic Hamiltonian of the radical pair system [12, 13] is expressed in Equation (1):

$$\hat{H} = \hat{H}_Z + \hat{H}_{HFI} + \hat{H}_J + \hat{H}_D \tag{1}$$

The spin Hamiltonian $\hat{H}$ comprises several key interactions: the Zeeman interactions of the two unpaired electron spins with the external magnetic field ($\hat{H}_Z$), the hyperfine interactions within the radical pair molecule ($\hat{H}_{HFI}$), and the inter-electron exchange ($\hat{H}_J$) and dipolar couplings ($\hat{H}_D$).

The Zeeman contribution to the spin Hamiltonian $\hat{H}_Z$ is given explicitly in Equation (2):

$$\hat{H}_Z(\theta, \phi) = \gamma_e B_0 \left( \hat{S}_x sin\theta cos\phi + \hat{S}_y sin\theta sin\phi + \hat{S}_z cos\theta \right) \tag{2}$$

Where $\gamma_e$ is gyromagnetic ratio of the electron ($\approx -28$ GHz/T), and $\hat{S}_x, \hat{S}_y$ and $\hat{S}_z$ are components of the electron spin operator in Cartesian coordinates. The direction of the external magnetic field is specified by the polar angles $\theta$ and $\phi$ in the molecular coordinate frame. Under geomagnetic field conditions (with a typical strength of $B_0 \approx 50 \, \mu T$), the difference and anisotropy in the g-factors of the two radicals can be neglected due to their similarity. As a result, the Zeeman interaction can be approximated using an isotropic form.

The hyperfine interaction component of the spin Hamiltonian, $\hat{H}_{HFI}$, is given in Equation (3):

$$\hat{H}_{HFI} = \sum_{I=1}^{2} \sum_{k=1}^{N} \left[ a^{(i,k)} \hat{\boldsymbol{S}}^{(i)} \cdot \hat{\boldsymbol{I}}^{(k)} + \hat{\boldsymbol{S}}^{(i)} \cdot \boldsymbol{A}^{(i,k)} \cdot \hat{\boldsymbol{I}}^{(k)} \right] \tag{3}$$

Here, $\hat{\boldsymbol{S}}^{(i)}$ and $\hat{\boldsymbol{I}}^{(k)}$ denote the electron and nuclear spin operators, respectively. The scalar term $a^{(i,k)}$ represents the isotropic (spherically symmetric) hyperfine coupling, while the tensor $\boldsymbol{A}^{(i,k)}$ describes the anisotropic contribution. In organic radicals, $\boldsymbol{A}^{(i,k)}$ is often approximated as axially symmetric, and its orientation in the molecular frame is defined by the Euler angles $\gamma(i,k)$ and $\delta(i,k)$[14].

Among the various interactions, hyperfine coupling and the geomagnetic field are usually the dominant factors controlling radical pair magnetosensitivity. However, when the inter-radical distance is short, both exchange and dipolar couplings become significant. The Hamiltonian $\hat{H}_J$ corresponding to the electron exchange interaction is given in Equation (4):

$$\hat{H}_J(r) = -J(r)\big(\hat{S}^2 - \hat{1}\big) \tag{4}$$

$$J(r) = J_0\,exp(-\beta r) \tag{5}$$

The exchange interaction $\hat{H}_J$ is governed by an exchange coupling constant $J(r)$, which is a function of the inter-radical distance $r$ (Equation (5)), and involves the total spin operator $\hat{S}^2$. This interaction leads to an energy splitting of $2J$ between the singlet $S$ and triplet $T_m$ states. The Hamiltonian $\hat{H}_D$ for the electron–electron dipolar interaction is described by Equations (6 and 7):

$$\hat{H}_D(r) = D(r)\left[\big(\hat{S}_z cos\varepsilon - \hat{S}_x sin\varepsilon\big)^2 - \frac{1}{3}\hat{S}^2\right] \tag{6}$$

$$D(r) = -\frac{3}{2}\frac{\mu_0}{4\pi}\frac{\gamma^2\hbar^2}{r^3} \tag{7}$$

which quantifies the dipolar coupling strength as a function of inter-radical distance $r$, with orientation specified in spherical coordinates.

Based on these theories, Schulten hypothesized that similar radical pair structures might form in the avian retina through either dark-state electron transfer or light-induced photoactivation. These radical pairs involve two unpaired electrons whose spin dynamics are influenced by anisotropic hyperfine interactions and external magnetic fields (Zeeman effect). The orientation of the external magnetic field alters the spin multiplicity of recombination products via back electron transfer (BET), providing a theoretical foundation for a biological compass. The reaction scheme for this radical pair system is conceptually summarized [4] in Equation (8):

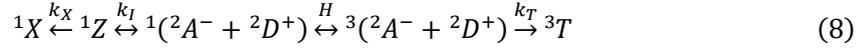

$$^{1}X \xleftarrow{k_X} {}^{1}Z \xleftrightarrow{k_I} {}^{1}(^{2}A^{-} + {}^{2}D^{+}) \xleftrightarrow{H} {}^{3}(^{2}A^{-} + {}^{2}D^{+}) \xrightarrow{k_T} {}^{3}T \qquad (8)$$

Here, $^{1}Z$ represents the singlet-state reaction precursor, corresponding to the singlet excited state $^{1}A^{*} + {}^{1}D$ generated under photoexcitation. Electron transfer leads to the formation of a radical pair in the $^{2}A^{-} + {}^{2}D^{+}$ state. The radical pair undergoes interconversion between singlet and triplet spin states, and subsequent spin-selective BET yields products in either the singlet $^{1}Z$ or triplet $^{3}T$ spin states. $^{1}X$ represents a signaling product that may be derived from the precursor $^{1}Z$ through a certain reaction pathway. The effect of the biological compass arises from the spin-state dynamics modulated by $\hat{H}$: as the orientation of the radical pair molecule with respect to the geomagnetic field varies (e.g., due to head movements in birds or other magnetosensitive organisms), the anisotropic nature of the hyperfine interaction leads to changes in the spin-state evolution. This, in turn, alters the ratio between the singlet $^{1}Z$ and triplet $^{3}T$ product yields. Consequently, changes in the magnetic field orientation can influence the yield of the signaling product $^{1}X$.

In the RPM-based model of biological magnetoreception, interconversion between electron spin states is primarily governed by the combined influence of hyperfine interactions and the external magnetic field (e.g., the geomagnetic field) via the Zeeman effect. In fixed-orientation molecular systems, hyperfine couplings exhibit pronounced anisotropy—i.e., the coupling strength depends on the orientation of the external magnetic field relative to the intrinsic molecular axes. This directional dependence leads to variations in the quantum coherent interconversion frequency between the singlet and triplet states as a function of the magnetic field orientation. Under the geomagnetic field (approximately 50 μT), although the Zeeman splitting is relatively weak, it is sufficient to lift the degeneracy of certain hyperfine states. As a result, the S–T interconversion dynamics become sensitive to the magnetic field direction, leading to a direction-dependent distribution of reaction products. This quantum dynamical behavior endows the radical pair with a "magnetic compass" function, enabling the organism to detect the direction of the geomagnetic field.

Whether the geomagnetic field can meaningfully influence the reaction of a radical pair in a spin-selective manner critically depends on the radical pair's lifetime, or more precisely, its spin coherence time. Within the framework of magnetoreception, the efficiency of singlet–triplet (S–T) interconversion is largely governed by the Larmor precession induced by the magnetic field. The Larmor time, which represents the timescale for a spin to complete one precessional cycle in a magnetic field, effectively sets a lower bound for magnetic field sensitivity. Thermal motion of surrounding nuclei introduces fluctuations in the hyperfine interactions, thereby perturbing the electron spins and inducing decoherence—this spin relaxation is a primary factor limiting the radical pair's functional lifetime. If the radical pair lifetime is shorter than the Larmor period, it cannot respond effectively to the magnetic field. Consequently, only when the lifetime exceeds the Larmor time can precessional motion significantly modulate spin dynamics, enabling the radical pair to exhibit magnetic field sensitivity. This spin-dynamic criterion is widely regarded as a fundamental physical prerequisite for biological magnetoreception, such as that hypothesized in avian navigation systems [15].

The geomagnetic field has a strength comparable to that of typical hyperfine interactions, allowing the RPM to effectively explain how external magnetic fields and hyperfine couplings influence the kinetics of radical recombination and the resulting product yields. In addition, the low-field effect (LFE) must be considered in radical pair systems under weak magnetic fields. LFE refers to the phenomenon in which, under low-intensity magnetic fields, slight perturbations to the hyperfine energy levels introduce additional resonance pathways that enhance singlet–triplet interconversion, ultimately increasing the yield of triplet products. This contrasts with high-field effects, wherein strong magnetic fields induce significant Zeeman splitting of the triplet sublevels, suppressing triplet formation [16]. The LFE is thus considered a mechanism that enhances the magnetic sensitivity of RPM-based magnetoreception. However, isotropic interactions such as electron exchange and electron–electron dipolar (EED) couplings, lacking angular dependence, disrupt the system's anisotropy and diminish the directional sensitivity to the magnetic field—posing a potential limitation to the "biological compass" function.

In the spin Hamiltonian defined in Equation (1), the terms $\hat{H}_J$ and $\hat{H}_D$ correspond to the electron exchange interaction and the EED interaction, respectively. The strengths of both interactions are highly sensitive to the distance separating the two unpaired electrons in the radical pair. In earlier theoretical studies, these terms were often neglected under the assumption that the radical pair constituents were sufficiently separated, rendering such interactions negligible. However, this assumption appears overly optimistic when applied to realistic biological systems.

For instance, in the widely studied magnetically sensitive radical pair $FAD^{\bullet-} - Trp^{\bullet+}$, the signal product is not derived from the singlet precursor (e.g., ${}^1Z$ in Equation (8)) but is instead formed directly from the radical pair itself. Consequently, product formation competes kinetically with radical recombination. To maintain spin selectivity, the signal yield must remain sensitive to spin-state evolution, implying that the timescale of BET should be comparable to that of signal formation. According to Marcus theory, achieving an optimal BET rate requires the edge-to-edge distance between the radicals to be approximately 1.5 nm [17]. At this separation, however, both exchange and dipolar interactions become non-negligible and may significantly influence the spin dynamics of the system. Under this dipolar interaction, the energy difference between the singlet state $S$ and the triplet states $T_m (m = 0, \pm 1)$ is given by $D\left(m^2 - \frac{2}{3}\right)$. Although both exchange and dipolar interactions cannot be neglected, their energetic effects can compensate or cancel each other if the condition $D = -3J$ or $D = 6J$ is met. This corresponds to an optimal inter-radical distance of approximately 2.0±0.2 nm [18].

The overall performance of a radical pair-based biological compass depends on several stringent physical and structural criteria: (a) Stability: The molecular axes defining the hyperfine interaction tensors must remain fixed relative to the entire radical pair system, necessitating spatially rigid or constrained molecular geometries within the biological environment; (b) Field-matching hyperfine coupling: The magnitude of hyperfine coupling should be comparable to the geomagnetic field (~50 µT) to enable magnetic field sensitivity; (c) Sufficient lifetime: The radical pair must persist longer than the Larmor precession period to allow sufficient spin-state evolution under magnetic influence; (d) Proper inter-radical distance: The radicals must be far enough apart to minimize disruptive exchange and dipolar interactions, but not so far as to hinder efficient back electron transfer [4, 19, 20].

In summary, an ideal radical pair for magnetoreception must exhibit long coherence times, spatial rigidity, appropriate hyperfine strength, and optimal inter-radical spacing. Maeda et al. demonstrated the feasibility of this model by synthesizing a carotenoid–porphyrin–fullerene system and observing anisotropic singlet product yields under magnetic field perturbations at cryogenic temperatures [21].

Therefore, RPM is well-suited to meet the navigational requirements of migratory birds, as it is sensitive primarily to the inclination of the geomagnetic field rather than its polarity. Additionally, two key parameters are commonly used in theoretical models to evaluate the angular sensitivity of the RPM-based system, in order to quantitatively assess the performance of such a magnetic compass.

The first key parameter used to evaluate magnetic field sensitivity in RPM-based models is the singlet quantum yield $\Phi_S$, defined as:

$$\Phi_S = k_S \int_0^\infty Tr\left(\widehat{P_S}\hat{\rho}(B,t)\right) dt \qquad (9)$$

where $k_S$ is the reaction rate constant for singlet recombination, $\widehat{P_S}$ is the projection operator onto the singlet state, and $\hat{\rho}(B,t)$ is the time-dependent spin density matrix governed by the stochastic Liouville equation. The orientation-averaged singlet yield, denoted $\overline{\Phi_S}$, is given by:

$$\overline{\Phi_S} = \frac{1}{2\pi} \int_0^{2\pi} \int_0^{\pi/2} sin(\theta)\, \Phi_S\big(B(\theta,\phi)\big) d\phi d\theta \tag{10}$$

where $B(\theta,\phi)$ represents the external magnetic field vector in spherical coordinates. The angular dependence of the singlet quantum yield $\Phi_S$ allows for the calculation of its average value $\overline{\Phi_S}$, as well as its maximum $\Phi_{S,\ max}$ and minimum $\Phi_{S,\ min}$ values. To quantify the directionality of the response, the singlet yield anisotropy $\Gamma_S$ is defined as:

$$\Gamma_S = \frac{\Phi_{S,\ max} - \Phi_{S,\ min}}{\overline{\Phi_S}} \tag{11}$$

which characterizes the magnetic directional sensitivity of the radical pair system. Another useful parameter is optimality (O), which reflects the spikiness of the response curve [22, 23], and is defined as:

$$O = \Gamma_S \left(1 - 2\frac{\overline{\Phi_S} - \Phi_{S,\ min}}{\Phi_{S,\ max} - \Phi_{S,\ min}}\right) \tag{12}$$

when the singlet yield varies smoothly with field direction (i.e., $\overline{\Phi_S} \approx \frac{1}{2}(\Phi_{S,\ max} + \Phi_{S,\ min})$), optimality approaches zero. In contrast, a highly spiked response—where enhanced yield occurs only in narrow angular regions—leads to a larger $O$, indicating sharper direction-selective behavior.

## 3. The Canonical FAD$^{\bullet-}$ − Trp$^{\bullet+}$ Radical Pair

Cryptochrome proteins are known to perform multiple biological functions, including the synchronization of circadian rhythms and the light-dependent regulation of plant growth and development [24, 25]. They are homologous to photolyases (DNA repair enzymes) [26], sharing a conserved photolyase homology region (PHR domain), while exhibiting highly variable N- and C-terminal extensions. The PHR domain binds a redox-active FAD cofactor in a non-covalent manner. When in its fully oxidized state, FAD is capable of absorbing blue light.

In 1993, Wiltschko et al. observed that magnetoreception behavior in birds depends on illumination at specific wavelengths (blue light), while birds exposed to red light lost their ability to navigate [27]. Based on this discovery, Ritz et al. hypothesized that cryptochrome proteins, which are responsive to blue light, may serve as key magnetoreceptors [5]. In addition to their well-known function in circadian regulation, cryptochromes have been shown in Drosophila melanogaster to modulate magnetic sensitivity in a blue-light-dependent manner: loss-of-function suppresses while overexpression enhances the response [28]. These findings collectively support the role of cryptochromes as light-dependent molecular mediators of magnetoreception in animals.

Although direct evidence is still lacking, light-manipulation experiments targeting Cluster N—a brain region implicated in compass-based navigation and part of the visual system in birds—provide indirect support [29, 30]. These findings suggest that cryptochromes localized in the retina, containing FAD cofactors capable of absorbing blue light, currently represent the most promising candidate magnetoreceptors.

Cryptochromes exist in multiple isoforms; for example, in the avian retina, several types are expressed, including Cry1, Cry2, and Cry4. However, a study examining the expression patterns of cryptochrome genes in the bird retina indicated that Cry4 is the most likely candidate involved in light-dependent magnetoreception, whereas Cry1 and Cry2 appear primarily associated with circadian rhythm regulation [31]. Subsequent in vitro experiments confirmed that Cry4 contains a magnetically sensitive radical pair, $FAD^{\bullet-} - Trp^{\bullet+}$, capable of detecting changes in magnetic field orientation [11]. Nonetheless, studies on other cryptochromes and homologous photolyases remain valuable for understanding the mechanisms by which Cry4 perceives magnetic fields.

Previous studies on other cryptochromes [32, 33] and photolyases [34] have revealed that the photoreduction of FAD is fundamentally driven by a sequential three-step electron transfer mediated by a conserved triad of tryptophan residues ( $Trp_A^{\bullet+}, Trp_B^{\bullet+}, Trp_C^{\bullet+}$ ), ultimately producing the flavin semiquinone radical ( $FAD^{\bullet-}$ ) (**Figure 1a**). Avian cryptochrome 4 (Cry4), however, exhibits a distinct feature, possessing a fourth tryptophan residue ($\text{Trp}_D^{\bullet+}$) involved in long-range electron transfer [35]. Experimental evidence confirms that the $\text{FAD}^{\bullet-} - \text{Trp}^{\bullet+}$ radical pair in Cry4 acts as a magnetosensitive radical pair [11]. This magnetic field sensing model is schematically illustrated in **Figure 1b**.

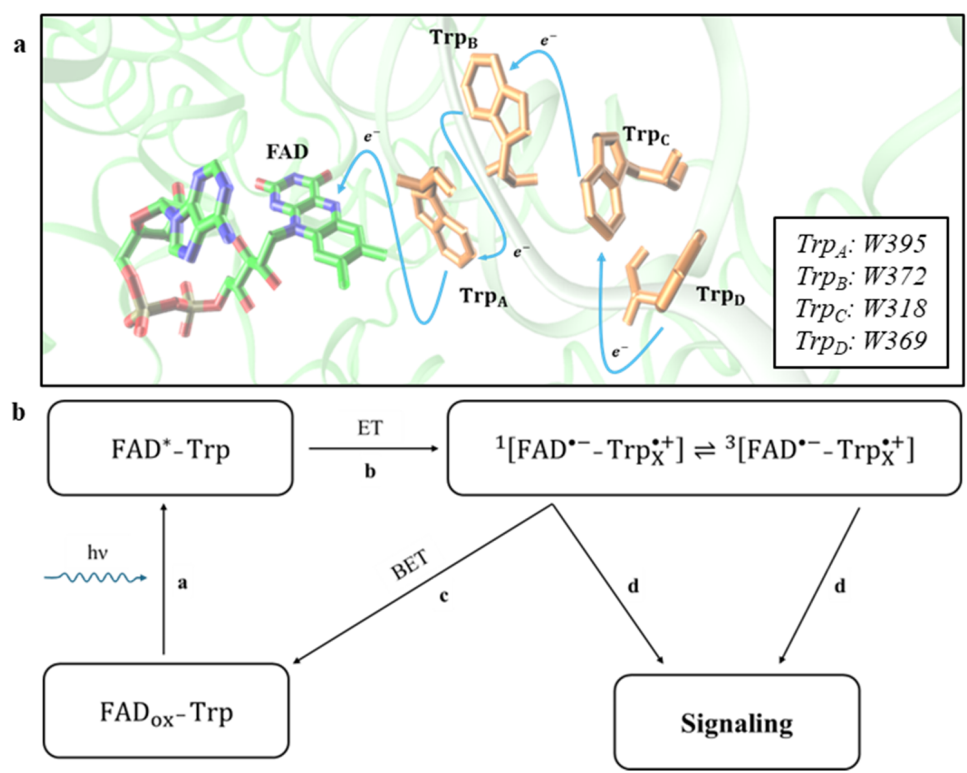

**Figure 1 (a) Schematic representation of the sequential electron transfer chain in avian cryptochrome 4 (Cry4).** *The conserved tryptophan triad ($\text{Trp}_A$, $\text{Trp}_B$, $\text{Trp}_C$) mediates stepwise photoreduction of the flavin adenine dinucleotide (FAD) cofactor, resulting in the formation of the flavin semiquinone radical ($\text{FAD}^{\bullet-}$). A fourth tryptophan residue ($\text{Trp}_D$) is uniquely present in Cry4 and facilitates long-range electron transfer. Arrows indicate the direction of electron flow;* **(b) $\text{FAD}^{\bullet-} - \text{Trp}_X^{\bullet+}$ Radical Pair Mechanism: Dependence on Tryptophan Position (X = A, B, C, D).** *Illustration of the $\text{FAD}^{\bullet-} - \text{Trp}_X^{\bullet+}$ radical pair mechanism, highlighting the effect of terminal tryptophan position (X = A, B, C, D) on spin dynamics and magnetoreception.*

The magnetic sensing process in cryptochrome proteins is initiated from the fully oxidized flavin cofactor ($\text{FAD}_{ox}$), a closed-shell molecule. Upon absorbing light in the 400–500 nm range (process a), an electron in $\text{FAD}_{ox}$ is promoted to a higher-energy orbital, yielding an excited singlet state $\text{FAD}^*$, in accordance with the Franck–Condon principle. Subsequently, an electron is transferred through a chain of conserved tryptophan residues

(process b), generating a key radical pair, $FAD^{\bullet-} - Trp_X^{\bullet+}$. Here, $Trp_X^{\bullet+}$ does not refer to a specific residue but rather denotes any tryptophan in the chain capable of forming a spin-selective radical pair. Given that the RPM is highly sensitive to perturbations such as exchange and dipolar interactions, the radical pair must maintain spin coherence. In avian Cry4, the distance between FAD and the terminal tryptophan residues ($Trp_C^{\bullet+}$ or $Trp_D^{\bullet+}$) is within the optimal range of $2.0 \pm 0.2$ nm, making them likely candidates for effective radical pair formation. The spin dynamics of the $FAD^{\bullet-} - Trp^{\bullet+}$ pair involve coherent interconversion between singlet and triplet states, modulated by hyperfine interactions (primarily from $^1H$ and $^{14}N$ nuclei), Zeeman interactions with the external magnetic field, exchange coupling, and dipolar interactions. This interconversion underlies the RPM's directional sensitivity to magnetic fields. The radical pair may undergo back electron transfer (BET) to regenerate the original $FAD_{ox}$ (process c), a spin-selective process restricted to the singlet state. In competition, a spin-independent pathway (process d) leads to the formation of $FADH^{\bullet}$, which is hypothesized to participate in downstream signaling processes essential for magnetic information transduction in birds.

In Arabidopsis thaliana, the signal transduction function of cryptochromes has been shown to correlate with conformational changes in the C-terminal region of the protein upon photoreduction of FAD to the semiquinone radical ($FADH^{\bullet}$) [15, 36-38]. A recent study on the animal-like cryptochrome CraCry from green algae revealed that large-scale structural rearrangements of the C-terminal $\alpha$-helix and the extended C-terminal tail of the protein are critically dependent on aspartic acid D321 [39]. D321 acts as a proton-coupled electron transfer (PCET) acceptor from tyrosine Y373, located near the terminal tryptophan of the electron transfer chain. Notably, a similar tyrosine residue is also present in avian cryptochromes such as ErCry4a, suggesting that analogous PCET processes may be involved in downstream magnetic signal transduction.

In summary, the presence of radical pairs within cryptochrome proteins provides a solid structural foundation for the RPM. The FAD cofactor, which efficiently absorbs blue light, offers a compelling explanation for the spectral dependence of avian magnetic navigation.

The rate of electron transfer (ET) between FAD and the conserved tryptophan residues along the electron transfer chain is a critical factor in the radical pair formation process (**Figure 1a**). The **Table 1** below summarizes the electron transfer rates—denoted as $ET_1$, $ET_2$, $ET_3$, $ET_4$—observed in cryptochromes and homologous photolyases. These correspond to the sequential electron transfer steps: $FAD \leftarrow Trp_A$, $Trp_A \leftarrow Trp_B$, $Trp_B \leftarrow Trp_C$, $Trp_C \leftarrow Trp_D$.

**Table 1 Tryptophan Chain Electron Transfer Rates in Cryptochromes and Photolyases** [a]

| Protein | $ET_1$ | $ET_2$ | $ET_3$ | $ET_4$ | Ref |
|---------|--------|--------|--------|--------|-----|
| AtCry1 | 0.4 ps | 4–15 ps | 30–50ps | - | [40-42] |
| DmCry | 0.85ps | 46ps | / | >3ns | [43] |
| ChCry4* | / | 23 ps (141 ps) | 149 ps (82 ps) | 57 ps (240 ps) | [44] |
| ErCry4a | 0.39 ps | 30 ps | 141 ps | 60ps* | [45] |
| EcPL | 0.8 ps | 70 ps (1.2 ns) | 150 ps (~100ns) | - | [46] |
| Xl (6-4) PL | 0.48 ps | 9.7 ps | 40 ps | 196 ps | [47] |

[a] Entries marked with an asterisk (*) represent data obtained from theoretical calculations. Values in parentheses indicate the rates of reverse electron transfer. A slash (/) denotes unavailable data, while a dash (–) indicates that the corresponding electron transfer step does not occur in the given protein.

Another key aspect of this reaction lies in the fact that spin recombination of the radical pair and the formation of spin-state-dependent signaling products are inherently competing processes. For the RPM to function as an effective magnetic sensor, both the back electron transfer (BET) and the formation of signaling products must occur on comparable time scales; otherwise, the signaling response would not effectively reflect external magnetic field changes. Whether the rate of BET can match that of signaling product formation largely depends on the distance between the terminal tryptophan residue and the FAD cofactor. In the X-ray crystal structure of ClCry4, the terminal tryptophan is positioned 1.76 nm and 2.13 nm away from FAD [48], which deviates from the theoretically optimal distance of approximately 1.5 nm. However, it has been suggested that a rapid dynamic equilibrium exists between $\text{Trp}_C^{\bullet+}$ and $\text{Trp}_D^{\bullet+}$, which minimizes the impact of distance on the BET rate. In fact, the $\text{FAD}^{\bullet-} - \text{Trp}_C^{\bullet+}$ pair, being more proximal, exhibits greater magnetic sensitivity and is theoretically better suited for magnetic field detection, whereas the $\text{FAD}^{\bullet-} - \text{Trp}_D^{\bullet+}$ pair, located closer to the protein surface, is more favorable for transmitting signals to the external environment. The fast interconversion between these two states balances their individual advantages, enhancing both the magnetic sensitivity and the efficiency of signal transduction in the cryptochrome system [35].

In addition to the intrinsic electron transfer rate, several internal factors—including environmental temperature, spin relaxation, and structural imperfections—also limit the performance of the $\text{FAD}^{\bullet-} - \text{Trp}_C^{\bullet+}$ radical pair as a magnetic compass. Experimental evidence has demonstrated that ErCry4 is capable of light-induced electron transfer and exhibits magnetic sensitivity under in vitro conditions; however, these measurements were conducted at 5 °C and pH ~8.0 [11]. In contrast, the intracellular environment of birds typically features a physiological temperature of 40–43 °C and a pH around 7.3 [49]. Although experimental conditions can be adjusted to match intracellular ionic states, the inevitable thermal motion of proteins at physiological temperature introduces complications. Enhanced thermal motion increases molecular vibrations, which in turn lead to fluctuations in hyperfine interactions—one of the primary "noise sources" disrupting radical pair coherence [8]. This disruption results in spin relaxation and loss of coherence.

A molecular dynamics (MD) study on Arabidopsis thaliana cryptochrome 1 (AtCry1) showed that elevated temperatures significantly impair the spin selectivity of $\text{FAD}^{\bullet-} - \text{Trp}^{\bullet+}$ radical pairs, rendering them virtually insensitive to Earth-strength magnetic fields [50].

However, as AtCry1 is of plant origin and is not under evolutionary pressure for magnetic field detection, it is plausible that avian cryptochromes have evolved a more favorable chemical environment to preserve the coherence of their radical pairs. In support of this, another MD simulation of avian Cry4 illustrated how internal molecular motions modulate electron–nuclear hyperfine couplings, and identified the frequency ranges of such motions that critically impact the directional sensitivity of RPM-based magnetic sensing [51]. Another theoretical study showed that in AtCry1 alone, the $FAD^{\bullet-} - Trp^{\bullet+}$ pair fails to maintain spin coherence for a sufficient Larmor precession time, implying that cryptochrome sensors may utilize structural constraints or binding interactions with cellular components to suppress unwanted degrees of freedom [50].

Notably, immobilization of the protein could reduce vibrational noise. MD simulations have suggested that Cry4 may be anchored at the membrane layers of the outer segments of double-cone photoreceptor cells—regions postulated to act as structural anchoring points [52-54]. This restricted orientation may be crucial for signal transduction and could enhance the magnetic field sensitivity of the radical pair by suppressing decoherence through limited molecular freedom.

Spin relaxation defects like those described above occur in most predicted radical pair systems, but the hyperfine structure limitations associated with the $FAD^{\bullet-} - Trp^{\bullet+}$ radical pair are unavoidable. Both radicals contain numerous nuclei exhibiting distinct, non-common symmetries in their hyperfine interactions, which significantly diminish magnetic sensitivity as the number of nuclei increases. As shown in **Figure 2a and 2b**, FAD and tryptophan possess 27 nuclei (including $^{14}N$ and $^{1}H$ ) that contribute to hyperfine interactions. To reduce computational cost, minor hyperfine couplings are often neglected, yet a substantial number of anisotropic hyperfine interactions remain. Key nuclei typically considered for FAD include N5, N10, $H\beta_1$, $H\beta_2$, H6, $H8_1$, $H8_2$, and $H8_3$, while for tryptophan, N1, H1, H2, H4, H5, H6, and H7 contribute to a complex hyperfine landscape. Under standard conditions, it is common to include only a subset of nuclei with the strongest hyperfine couplings to the radical electrons [53]. Nonetheless, increasing the number of nuclei considered invariably leads to a pronounced decrease or even loss of radical pair magnetic sensitivity due to enhanced decoherence induced by thermal motion-driven noise from these hyperfine interactions.

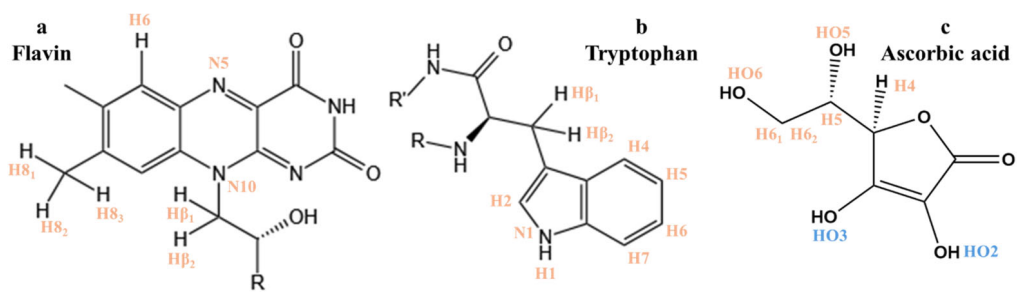

**Figure 2** (a) The flavin moiety of FAD, with R representing the remainder of the ribose, phosphate linker, and AMP groups; (b) The tryptophan residue in cryptochrome, where R and R′ denote the peptide chains flanking the residue; (c) Vitamin C (AscH$_2$) undergoes deprotonation at the hydroxyl groups at positions 2 and 3, forming the ascorbate radical (Asc$^{\bullet-}$).

Moreover, when the partner radical in the pair lacks hyperfine structure, the singlet–triplet (S–T) interconversion frequency is governed primarily by magnetically isolated electrons interacting solely with the geomagnetic field, resulting in S–T oscillations at the Larmor frequency (~1.4 MHz). However, when the partner radical contains hyperfine structure, increasing hyperfine complexity rapidly dominates this frequency, broadening and washing out distinct spectral features; the S–T interconversion frequency distribution can span from 0.1 to 12 MHz [15]. Interestingly, this broader frequency distribution may enhance S–T interconversion efficiency for radical pairs with lifetimes on the order of 1 ps, since hyperfine-induced frequency broadening effectively raises the overall oscillation frequency, providing some gain in magnetic sensitivity despite the aforementioned decoherence "noise" effects.

Additionally, the key performance metrics of the avian "magnetic compass," namely the singlet yield anisotropy and optimality, have been shown to be tunable by adjusting the spatial orientation of the FAD$^{\bullet-}$ − Trp$^{\bullet+}$ radical pair. Modulation of the molecular geometry can influence the hyperfine interactions within the system, allowing both metrics to approach their optimal values. However, the parameter sets required to optimize each metric differ substantially, resulting in an inherent trade-off and a negative correlation between singlet yield anisotropy and optimality. Notably, naturally occurring cryptochrome proteins appear to have evolved to achieve a balanced, functionally viable compromise between these competing factors [23].

Finally, a major limitation of the $FAD^{\bullet-} - Trp^{\bullet+}$ radical pair lies in its inability to effectively cancel the exchange interaction and the dipolar interaction, with the latter being non-negligible. Experimental quantification comparing the exchange interaction ($J_{ex}$) and dipolar interaction (D) in Drosophila cryptochrome (DmCry) and E. coli photolyase (EcPL) revealed that the exchange interaction strength is at least one order of magnitude smaller than that of the dipolar interaction, thus precluding effective cancellation between them. Specifically, the exchange interaction in DmCry, approximately 0.43 MHz, can be neglected in radical pair studies, whereas the dipolar interaction, around 4 MHz, is sufficiently strong to significantly affect hyperfine coupling [55]. Theoretical studies on DmCry further indicate that when dipolar interactions are included, the electron dipolar and exchange interactions are difficult to counterbalance, and the magnetic compass sensitivity is diminished by the dipolar interaction [56].

External oscillating magnetic fields have been shown to interfere with avian magnetoreception, providing compelling support for the RPM. For instance, studies on the European garden warbler demonstrated that cryptochromes containing millisecond-lived tryptophan-derived radical pairs are disrupted by a weak oscillating magnetic field at 1.4 MHz and 190 nT, impairing the bird's geomagnetic orientation [57, 58]. This type of disruption has been recognized as a key line of evidence supporting the RPM [59-63].

In organic radical systems, the ability to respond to additional weak oscillating fields (particularly in the MHz range) while being sensitive to static magnetic fields below 1 mT has been proposed as a diagnostic test for the presence of an RPM [64]. The principle behind this test lies in the resonance between the singlet–triplet (S–T) interconversion frequency— determined by Zeeman and hyperfine interactions—and the applied radiofrequency field. When the external field matches the intrinsic oscillation frequency of the radical pair system, resonant enhancement of S–T mixing occurs, resulting in measurable changes in reaction yields. This phenomenon is commonly referred to as reaction yield detected magnetic resonance (RYDMR) and allows for inference of the radical pair's internal frequencies, as well as estimates of hyperfine and Zeeman coupling strengths.

Although the real $FAD^{\bullet -} - Trp^{\bullet +}$ radical pair features numerous symmetry-breaking hyperfine couplings, which would be expected to lead to a broad and frequency-averaged response to RF fields, this is not always the case. The Joseph S. Takahashi lab reported that a radio filed (RF) field at the Larmor frequency (1.315 MHz) with an amplitude as low as 15 nT could effectively disrupt magnetic orientation in birds [48]. In contrast, fields at half (0.65 MHz) or double (2.63 MHz) the frequency required more than 300-fold higher amplitudes to induce similar effects. Such a sharply tuned frequency sensitivity is inconsistent with expectations for a radical like $Trp^{\bullet +}$, which is rich in hyperfine structure. This observation instead suggests the possible involvement of a secondary radical species ($Z^{\bullet}$) lacking significant hyperfine interactions.

## 4. Radical pairs involving $O_2^{\bullet -}$

There is little doubt that the RPM, once experimentally validated, is both realistic and feasible [21]. However, the actual radical pair components operating in vivo remain a subject of debate. Within cryptochromes, FAD is widely regarded as the indispensable chromophore responsible for magnetic sensitivity. The canonical radical pair, $FAD^{\bullet -} - Trp^{\bullet +}$, has been the most extensively studied model. Yet, several unresolved issues persist. Chief among them is the inability of this pair to explain why a weak radiofrequency field at the Larmor frequency significantly disrupts avian magnetoreception. In addition, the anisotropic performance of this radical pair is substantially diminished due to the complex hyperfine interactions associated with both radicals, limiting its functional relevance under physiological conditions. These shortcomings suggest that a more plausible radical pair in vivo might involve $FAD^{\bullet -}$ coupled with a second radical species lacking significant hyperfine interactions. In fact, several studies propose that the absence of hyperfine-active nuclei in the secondary radical is not only advantageous but may be essential for achieving the levels of magnetic field sensitivity observed in nature [60].

In biological systems, it is conceivable that a radical species $Z^{\bullet}$, minimally affected by hyperfine-active nuclei (such as $^1H$ and $^{14}N$), could be involved in magnetoreception. Following the formation of the conventional $FAD^{\bullet-} - Trp^{\bullet+}$ radical pair via electron transfer along the conserved tryptophan chain, the terminal tryptophan residue may undergo electron transfer with a surface-accessible molecule Z, giving rise to a new radical pair: $FAD^{\bullet-} - Z^{\bullet}$. Theoretically, due to the asymmetry in hyperfine interactions, this alternative radical pair has been predicted to exhibit magnetic directional sensitivity to the geomagnetic field that is up to two orders of magnitude greater than that of the classical $FAD^{\bullet-} - Trp^{\bullet+}$ system [10]. A study on Arabidopsis thaliana cryptochrome (AtCry1) compared spin relaxation effects in $FAD^{\bullet-} - Trp^{\bullet+}$ and the putative $FAD^{\bullet-} - Z^{\bullet}$ pair under thermal molecular motion. The results suggest that the latter is less susceptible to decoherence induced by molecular vibrations, thereby potentially affording a longer spin-coherence lifetime and enhanced magnetic sensitivity [50]. Among the candidates proposed for the secondary radical $Z^{\bullet}$, two species have been most frequently suggested: the superoxide anion radical and the ascorbate radical.

Compared to $O_2^{\bullet-}$, the ascorbate radical offers a key advantage in that it does not undergo rapid spin relaxation. The unpaired electron in $Asc^{\bullet-}$ is primarily delocalized over the enediol moiety between the C2 and C3 positions and the adjacent oxygen atoms, with significant hyperfine coupling only involving the nearby H4 nucleus (**Figure 2c**). This limited hyperfine interaction makes $Asc^{\bullet-}$ a viable candidate for forming magnetically sensitive radical pairs. Under acidic aqueous conditions, ascorbate can transfer an electron to photoexcited flavin mononucleotide (FMN) to generate radical pairs that are magnetically responsive under weak magnetic fields as low as 0.55 mT [65]. Binding studies of $Asc^{\bullet-}$ with DmCry and European robin cryptochrome 1a (ErCry1a) indicate that its residence time near the tryptophan radical site is on the nanosecond timescale. However, in physiological conditions, the intracellular concentration of ascorbate is typically below 1 mM. Molecular dynamics simulations suggest that this low concentration may limit its effective participation in magnetosensory processes [66].

In addition to serving as a possible replacement for the tryptophan radical in the canonical $FAD^{\bullet-} - Trp^{\bullet+}$ pair, the radical $O_2^{\bullet-}$ has also been proposed as a potential partner in magnetoreception involving cryptochromes under dark-state conditions. Upon photoactivation, cryptochromes undergo a redox cycle comprising a light-dependent photoreduction step and a light-independent reoxidation process. The former leads to the formation of a radical pair in the form of $FAD^{\bullet-} - Trp^{\bullet+}$, while the latter is hypothesized to involve an electron transfer reaction between $FADH^-$ and molecular oxygen ($O_2$), yielding an alternative radical pair: $FADH^{\bullet} - O_2^{\bullet-}$. Rather than replacing the $FAD^{\bullet-} - Trp^{\bullet+}$ radical pair, the potential involvement of $O_2^{\bullet-}$ in forming a magnetically sensitive pair with $FADH^{\bullet}$ may complement the canonical mechanism. These two pathways could operate synergistically, each playing a role at different steps of the cryptochrome redox cycle in facilitating magnetic sensing.

In fact, cryptochrome-mediated magnetic field sensing does not rely exclusively on continuous light exposure; it can also occur under dark conditions following intermittent illumination. Behavioral experiments on birds employing alternating light and magnetic pulses have demonstrated that birds remain responsive to magnetic fields during dark intervals between light exposures. This finding suggests that radical pairs involved in geomagnetic sensing may arise during the reoxidation process of FAD [67]. Similarly, intermittent light experiments in Arabidopsis thaliana showed that cryptochrome-dependent magnetosensitivity persists in the dark and is independent of light-induced flavin electron transfer [68, 69].

Targeted photoactivation of chicken cryptochrome Cry1a with wavelengths ranging from 400 to 565 nm revealed activation beyond the visible absorption spectrum of FAD (400–500 nm), indicating the presence of additional chromophores within the compass system.

This extended absorption has been attributed to the FADH$^\bullet$ radical [70, 71]. Experimental studies of photoreduction and reoxidation of plant cryptochrome AtCry1 confirmed that molecular oxygen plays a critical role as an oxidant in flavin reoxidation. The proposed reaction pathway proceeds via photoreduction of $FAD_{OX}$ to $FAD^{\bullet-}$, followed by protonation to form FADH$^\bullet$, and further reduction to $FADH^-$, which then reacts with oxygen to form the $FADH^\bullet - O_2^{\bullet-}$ radical ion pair, ultimately reoxidizing to $FAD_{OX}$ [73].

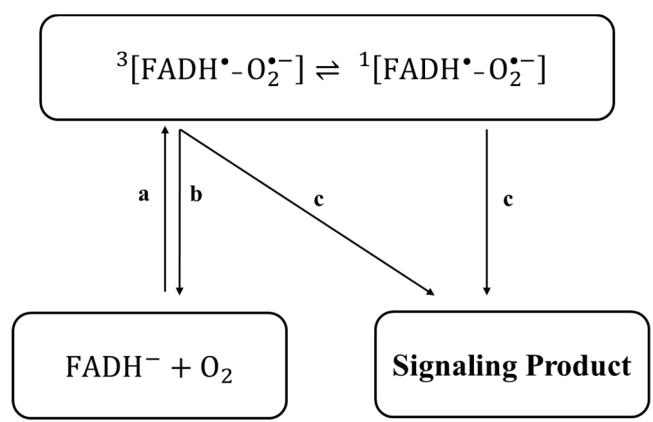

**Figure 3 Proposed RPM mechanism involving the FADH$^\bullet$ − $O_2^{\bullet-}$ system.** *The scheme illustrates the formation and spin-selective recombination pathways of the $FADH^\bullet - O_2^{\bullet-}$ radical pair within the cryptochrome protein under geomagnetic field influence.*

Based on this pathway, cryptochrome hosts an RPM involving the $FADH^\bullet - O_2^{\bullet-}$ system as illustrated in **Figure 3** The photoreduction product $FADH^-$ undergoes electron transfer via pathway a, generating a triplet state $FADH^\bullet - O_2^{\bullet-}$ radical pair. Under anisotropic hyperfine interactions, singlet-triplet (S–T) interconversion occurs. Analogous to the $FAD^{\bullet-} - Trp^{\bullet+}$ system, the downstream formation of signaling products via pathway c is spin-independent, whereas only the triplet state can regenerate $FADH^-$ and oxygen via back electron transfer (pathway b). For example, when the magnetic field orientation changes from parallel to perpendicular relative to the flavin molecular plane, the proportion of triplet products formed during S–T interconversion increases while the signaling product yield decreases. The spin-selective modulation of S–T interconversion and pathways b and c by magnetic field direction constitutes the core mechanism of the $FADH^\bullet - O_2^{\bullet-}$ radical pair RPM.

Regardless of whether the superoxide anion forms a radical pair in the form of $FAD^{\bullet-} - O_2^{\bullet-}$ or $FADH^\bullet - O_2^{\bullet-}$, both configurations enable the cryptochrome protein system to meet the requirement of asymmetric hyperfine coupling, which is critical for directional magnetic sensing. Theoretically, such configurations are predicted to yield high singlet yield anisotropy ($\Gamma_S$), and are consistent with experimental observations demonstrating sensitivity to radiofrequency magnetic fields. Furthermore, $O_2^{\bullet-}$ radicals have been shown to participate in magnetically sensitive reactions; for example, in $NO - O_2^{\bullet-}$ systems, magnetic field effects have been observed to alter recombination pathways, indicating that superoxide possesses the fundamental prerequisites for involvement in magnetic field-sensitive radical pair reactions [72]. However, radical pairs containing $O_2^{\bullet-}$ also exhibit notable limitations. Most critically, the superoxide anion radical undergoes extremely rapid spin relaxation, due to strong coupling between its electron spin and the molecular principal axis. This causes the electron spin state to be highly sensitive to molecular vibrations and rotations, leading to loss of spin coherence between the radicals [74, 75].

Molecular dynamics simulations and electronic structure calculations further suggest that, in DmCry, superoxide can diffuse into the flavin-binding pocket and form either a $FAD^{\bullet-} - O_2^{\bullet-}$ or protonated $FAD^{\bullet-} - HO_2^\bullet$ complex. These radical pairs may be stabilized via hydrogen bonding, which restricts $O_2^{\bullet-}$ rotational freedom. However, the short distance between the flavin ring and $O_2^{\bullet-}$ (~0.3 nm) induces a very strong exchange interaction—estimated at ~345 T, approximately 7 million times stronger than Earth's magnetic field—which is sufficient to completely suppress the functionality of the radical pair [75, 76]. Nonetheless, some studies have proposed that if recombination processes exhibit strong asymmetry, quantum Zeno effects may enable even tightly bound radical pairs to retain sensitivity to weak geomagnetic fields [77, 78].

Considering the specific requirements for radical pairs involving $FAD^{\bullet-} - Trp^{\bullet+}$ and the physicochemical properties of $O_2^{\bullet-}$, the radical pairs $FAD^{\bullet-} - O_2^{\bullet-}$ or $FADH^{\bullet} - O_2^{\bullet-}$ must fulfill the following criteria to function effectively in a magnetoreceptive system:

Suppression of molecular motion: To prevent rapid spin relaxation, the $O_2^{\bullet-}$ radical must be restricted from undergoing significant rotational or translational movement. This may be achieved if $O_2^{\bullet-}$ binds to a specific site within the cryptochrome protein, thereby creating a microenvironment that stabilizes its spin coherence.

Optimal radical pair separation: The distance between $O_2^{\bullet-}$ and the flavin moiety should be within the range of 1.5–2.0 nm to allow for efficient radical recombination. This spatial arrangement also minimizes undesirable exchange and dipolar interactions, which are strongly distance-dependent.

Matching reaction kinetics: Assuming spin relaxation of $O_2^{\bullet-}$ is suppressed, the rate of singlet recombination—converting $FADH^{\bullet} - O_2^{\bullet-}$ back to $FADH^-$ and molecular oxygen, and subsequently reoxidizing to $FAD_{OX}$—must occur on a similar timescale as the S–T (singlet–triplet) interconversion, without premature spin decoherence.

Avoidance of spin–orbit coupling: For spin-selectivity to be retained, the energy gap between the two $\pi^*$ orbitals of $O_2^{\bullet-}$ must exceed its spin–orbit coupling constant ($\approx 200$ cm$^{-1}$). This condition implies that $O_2^{\bullet-}$ must exist in a highly asymmetric local electronic environment to suppress spin–orbit-induced relaxation.

At present, research on radical pairs involving $O_2^{\bullet-}$ remains limited. On the one hand, there is insufficient direct experimental evidence demonstrating that molecular oxygen can actively participate in radical pair mechanisms. On the other hand, it remains unclear whether radical pairs involving $O_2^{\bullet-}$ can satisfy the stringent physicochemical criteria outlined above.

First, there is still a lack of theoretical and experimental support for the notion that cryptochrome proteins can bind $O_2^{\bullet-}$ effectively; the specific binding sites and modes of interaction have not yet been identified. Second, according to equations (7) and (8), the dipolar interaction in a rigid system primarily depends on the distance between the unpaired electrons and their relative spin orientations. This interaction is an intrinsic feature of the magnetosensory system and does not depend on external environmental factors. In fact, at a typical distance of ~2.0 nm, the dipolar interaction strength is estimated to be ~0.36 mT [79], which is on the same order of magnitude as typical hyperfine couplings (~1 mT). Moreover, this dipolar interaction cannot be fully canceled out by exchange interactions. In systems involving either $FAD^{\bullet-} - Trp^{\bullet+}$ or $FAD^{\bullet-} - Z^{\bullet}$ radical pairs, the introduction of EED interactions cause the anisotropic singlet yield ($\Gamma_S$) to converge across different magnetic field orientations, substantially diminishing the system's sensitivity to magnetic field direction. These findings suggest that the feasibility of such mechanisms remains debatable [56].

Current research suggests that introducing a suitable third radical into a traditional radical pair system to form a triad $[A^{\bullet} - B^{\bullet} - C^{\bullet}]$ may mitigate the detrimental effects caused by EED interactions. The third radical can potentially alleviate EED-induced issues through two major mechanisms. The first is known as the scavenging mechanism, which involves a spin-selective electron transfer reaction between one member of the original radical pair and a paramagnetic scavenger. In this scheme, spin-selective recombination of the original singlet or triplet radical pair via back electron transfer is not essential [80].

One proposed scavenging mechanism is illustrated in **Figure 4**: in pigeon cryptochrome ClCry4, tyrosine Y319 acts as a scavenger molecule, participating in the magnetosensitive $FADH^{\bullet} - O_2^{\bullet-}$ radical pair pathway. Upon photo-reduction, $FADH^-$ undergoes an electron transfer (pathway a) to generate the radical pair $FADH^{\bullet} - O_2^{\bullet-}$. Due to the relatively long separation between the radicals, back electron transfer (pathway b) is negligible, and the pair instead yields signaling products via pathway c. Tyrosine Y319 interacts with $FADH^{\bullet}$ through a spin-selective electron transfer chain involving neighboring tryptophan residues, thus allowing the system to transduce magnetic field effects [81].

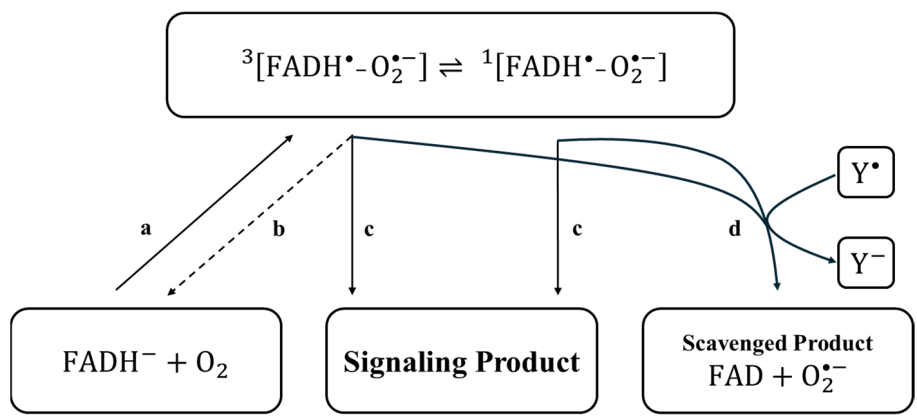

**Figure 4** Tyrosine Y319 Involvement in the Scavenging Mechanism of the $FADH^{\bullet} - O_2^{\bullet-}$ Radical Pair. This scheme outlines the Y319-dependent scavenging pathway and provides a plausible basis for the feasibility of an RPM involving a superoxide-containing radical pair.

The advantages of the scavenger mechanism are twofold: (a) It permits the radicals within the magnetoreceptor protein to be spaced more than 2 nm apart, significantly reducing the negative influence of exchange and dipolar interactions, thereby enhancing the anisotropy of the reaction [82]; (b) If one radical in the pair undergoes rapid spin relaxation, this can paradoxically increase the spin-dependent anisotropy of the reaction [83];

The second strategy involves the bystander radical mechanism, in which a third radical species ($B^{\bullet}$) is positioned either within or on the surface of the cryptochrome protein. When appropriately located—for instance, near the terminal tryptophan residue on the protein surface—this bystander radical can modulate the exchange interaction with $Trp^{\bullet+}$ or $Z^{\bullet}$. By fine-tuning the exchange interaction strengths, the bystander can reduce the EED (electron–electron dipolar) interaction between the flavin and the other radical, resulting in a pronounced peak in the singlet anisotropy yield $\Gamma_S$ under varying magnetic field orientations [56]. In addition to externally sourced radicals, tyrosine residues located near the terminal tryptophan may serve as endogenous radical candidates due to their relatively long lifetimes, making them particularly promising bystander radical $B^{\bullet}$ species. Moreover, the two proposed mechanisms are not mutually exclusive; during the scavenging process, the third radical may still function as a bystander radical, potentially contributing to the overall magnetic sensitivity of the system

## 5. Conclusion and Outlook

This review summarizes the fundamental principles of the radical pair mechanism in cryptochrome proteins, with a particular focus on the magnetic properties and limitations of the canonical $FAD^{\bullet-} - Trp^{\bullet+}$ pair and alternative radical pairs involving superoxide. To facilitate a clear comparison of different radical pairs, we provide here a concise summary of the principal radical species and their key physical parameters (see **Table 2**). This comparison not only emphasizes the distinctions in physical properties between the canonical $FAD^{\bullet-} - Trp^{\bullet+}$ pair and possible alternative radical pairs (e.g., $FADH^{\bullet} - O_2^{\bullet-}$), but also serves as a reference for guiding future experimental investigations and theoretical simulations.

**Table 2 Comparative Summary of the Physical Properties of Radical Pairs[a]**

| RP | Hyperfine interactions included | Distance (nm) | Lifetime | Ref |
|---|---|---|---|---|
| $FAD^{\bullet-} - Trp_C^{\bullet+}$ | >20 | 1.76 | 1~10 µs | 11, 48 |
| $FAD^{\bullet-} - Trp_D^{\bullet+}$ | >20 | 2.13 | 1~10 µs | 11, 48 |
| $FADH^{\bullet} - O_2^{\bullet-}$ | 2~3 | 0.3~0.4 | n.d. (expected < ns) | 15, 76 |
| $FADH^{\bullet} - Y^{\bullet}(O_2^{\bullet-})^{[b]}$ | 2~3 | 1.71 | n.d. (depends on $Tyr^{\bullet}$) | 15, 81 |

[a] The radical pairs listed in the table are all considered within ClCry4. The physical parameters include the number of hyperfine interactions accounted for, the inter-radical distance, and the radical pair lifetime. [b] In the case of $FADH^{\bullet} - Y^{\bullet}(O_2^{\bullet-})$, this denotes the relationship between flavin and tyrosine within the radical triad system.

Although direct experimental evidence for the involvement of $O_2^{\bullet-}$ in the cryptochrome radical pair mechanism remains limited, indirect behavioral studies—such as the disruption of avian magnetoreception by radiofrequency magnetic fields—strongly suggest that $O_2^{\bullet-}$ plays a critical role in magnetic sensing.

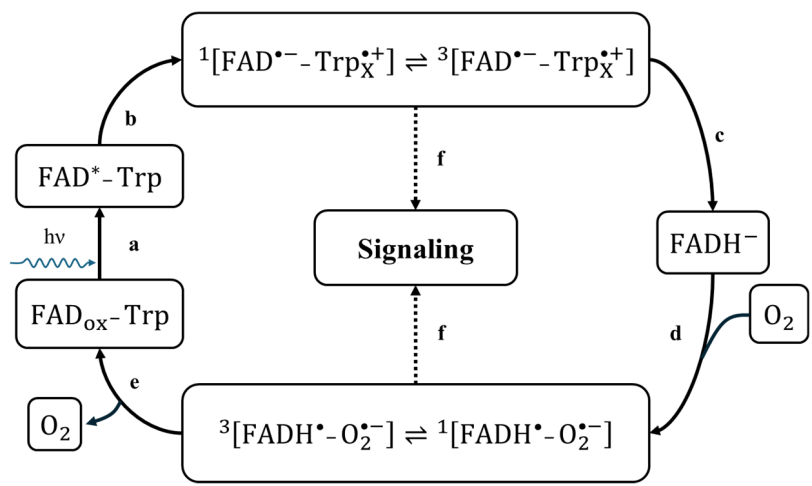

**Figure 5 (a)** Upon absorption of blue light, $FAD_{OX}$ is excited to its singlet excited state $FAD^*$; (b) A subsequent photoinduced electron transfer leads to the formation of a radical pair, $FAD^{\bullet-} - Trp_X^{\bullet+}$; (c) $FAD^{\bullet-}$ is further reduced via photoreduction and proton transfer to yield $FADH^-$; (d) FFADH$^-$ donates an electron to molecular oxygen, producing the radical pair $FADH^{\bullet} - O_2^{\bullet-}$; (e) In the reoxidation step, $FADH^{\bullet}$ and $O_2^{\bullet-}$ recombine to regenerate the oxidized flavin $FAD_{OX}$; (f) Signal transduction is proposed to occur during or following the radical pair processes, potentially forming the molecular basis for magnetoreception.

If $O_2^{\bullet-}$ participates in the reoxidation step of the cryptochrome photocycle, forming a radical pair with $FADH^{\bullet}$, this process could complement the $FAD^{\bullet-} - Trp^{\bullet+}$ reaction and collectively constitute a redox cycle that underpins magnetosensory function. A schematic representation of this cycle is shown in **Figure 5** [81]. Within this cycle, the $FADH^{\bullet} - O_2^{\bullet-}$ pair may interact with an additional third radical, forming a radical triad system that enhances magnetic sensitivity through multispin interactions.

Since the pioneering work of Wiltschko and Merkel in 1966 [84], which demonstrated that birds can navigate using the geomagnetic field, the molecular mechanisms underlying avian magnetoreception have remained elusive despite over five decades of intensive investigation. In particular, the radical pair mechanism proposed to operate in cryptochrome proteins is still far from being fully understood. Experimentally, it remains unclear whether cryptochromes can retain magnetic sensitivity under the physiologically relevant, thermally noisy, and aqueous intracellular conditions of living cells. The pathways by which magnetically sensitive signals are generated, amplified, and transduced are also unknown. To date, no high-resolution structure of cryptochromes from night-migratory birds has been reported. While the sequential electron transfer along the tryptophan triad, leading to the formation of the canonical $FAD^{\bullet -} - Trp^{\bullet +}$ radical pair, is widely accepted, the possibility of molecular oxygen acting as a spin partner in alternative radical pairs remains unverified experimentally.

Furthermore, the deleterious effects of EED and exchange interactions on the magnetic sensitivity of radical pairs cannot be ignored, yet it remains uncertain how such detrimental influences are circumvented by nature. For theoretically proposed magnetically sensitive species such as $FADH^{\bullet} - O_2^{\bullet -}$, how $O_2^{\bullet -}$ is stabilized within the cryptochrome scaffold to prevent rapid rotational diffusion—a process that would otherwise quench magnetosensitivity—is a critical open question.

Recent advances in molecular dynamics and spin dynamics simulations have enabled detailed theoretical investigations into how cryptochrome-based magnetoreception models respond to changes in magnetic fields and transmit relevant signals. These computational approaches represent a leading direction in the field. However, the theoretical predictions still require decisive experimental validation to establish their reliability. Moreover, current models remain incomplete. The adverse effects of EED and exchange interactions on magnetic sensitivity, the mechanisms by which electron transfer is achieved within cryptochromes of migratory birds, and how these proteins preserve functional magnetic sensing under the thermally noisy and complex intracellular environment all remain open questions. Addressing these challenges will require simulations of biologically relevant cryptochrome structures and their native cellular contexts. Although remarkable advances have been made in elucidating the physical chemistry of magnetoreception over the past fifty years, many fundamental scientific questions remain unanswered.

Prospectively, several promising research directions, spanning molecular, physicochemical, computational, and behavioral levels, may lead to breakthroughs in addressing the current challenges in cryptochrome-mediated magnetoreception. At the molecular level, high-resolution structural determination remains essential; however, low protein abundance and conformational heterogeneity significantly limit experimental access, and while homologous or AI-predicted models provide partial guidance, they cannot fully replace native structures. At the physicochemical level, techniques such as ultrafast spectroscopy, EPR, and NMR are indispensable for monitoring protein dynamics, radical pair formation, spin coherence, and hyperfine interactions, thereby providing stringent experimental validation for theoretical predictions. Computationally, multiscale simulations that integrate electronic structure, molecular dynamics, and spin dynamics can elucidate the principles of radical pair mechanisms, the behavior of alternative radical pairs, electron-transfer-induced conformational changes (particularly in the C-terminal region of cryptochromes), and the transduction of magnetic signals. Finally, combining biochemical, genetic, and in vivo behavioral studies is critical for linking molecular-level mechanisms to birds' magnetic navigation. Collectively, these interdisciplinary efforts will pave the way toward a more comprehensive and experimentally grounded understanding of radical pair mechanisms and provide a clear roadmap for future investigations into the molecular basis of avian magnetoreception. It is our hope that this concise overview of the radical pair mechanism in avian magnetoreception will stimulate further research and attract the attention of scientists across disciplines to this intellectually challenging and biologically significant field.

## Acknowledgments

This work was supported by the Natural Science Foundation of Guangdong Province (No. 2024A1515010304) and the National Natural Science Foundation of China (No. 31971176).